\documentclass{cernrep}
\newcommand{\be}{\begin{equation}}
\newcommand{\ee}{\end{equation}}

\newcommand{\Tr}{{\rm Tr}}
\begin{document}
\title{Microsopic nuclear level densities by the shell model Monte Carlo method}
\author{Y. Alhassid$^1$, G.F.~Bertsch$^{2}$, C.N.~Gilbreth$^{1}$,  H. Nakada$^3$, and C. \"Ozen$^4$}

\institute{$^{1}$Center for Theoretical Physics, Sloane Physics 
Laboratory, Yale University, New Haven, Connecticut 06520, USA\\ 
$^{2}$Department of Physics and Institute for Nuclear Theory, 
Box 351560, University of Washington, Seattle, Washington 98915, USA\\
$^3$Department of Physics, Graduate School of Science, Chiba University, Inage, Chiba 263-8522, Japan\\
$^4$Faculty of Engineering and Natural Sciences, Kadir Has University, Istanbul 34083, Turkey}

\maketitle

\begin{abstract}
The configuration-interaction shell model approach provides an attractive framework for the calculation of nuclear level densities in the presence of correlations, but the large dimensionality of the model space has hindered its application in mid-mass and heavy nuclei.  The shell model Monte Carlo (SMMC) method permits calculations in model spaces that are many orders of magnitude larger than spaces that can be treated by conventional diagonalization methods. We discuss recent progress in the SMMC approach to level densities, and in particular the calculation of level densities in  heavy nuclei. We calculate the distribution of the axial quadrupole operator in the laboratory frame at finite temperature and demonstrate that it is a model-independent signature of deformation in the rotational invariant framework of the shell model. We propose a method to use these distributions for calculating level densities as a function of intrinsic deformation.  
\end{abstract}

\section{Introduction}

Nuclear level densities are an integral part of the calculation of transition rates through Fermi's Golden rule and of the Hauser-Feshbach theory~\cite{Hauser1952} of statistical nuclear reactions. However, their microscopic calculation the presence of correlations is a challenging many-body problem.  Theoretical models of level density are often based on mean-field and combinatorial methods~\cite{Konig2008} but they can miss important correlations. 

The configuration-interaction (CI) shell model approach accounts for correlations and shell effects, but conventional diagonalization methods are limited to spaces of dimensionality $\sim 10^{11}$. The shell model Monte Carlo (SMMC) method~\cite{Lang1993,Alhassid1994,Koonin1997,Alhassid2001} permits calculations in model spaces that are many orders of magnitude larger than those that can be treated by conventional methods.  Quantum Monte Carlo methods for fermions often have a sign problem that limits their applicability. However, the dominant collective components~\cite{Zuker1996} of effective nuclear interactions have a good Monte Carlo sign in SMMC and are sufficient for realistic calculation of level densities and collective properties of nuclei. Small bad-sign components of the nuclear interaction can be treated in the method of Ref.~\cite{Alhassid1994}. 

As a finite-temperature method, SMMC is particularly suitable for the calculation of state densities~\cite{Nakada1997}. It has been applied successfully to mid-mass nuclei~\cite{Alhassid1999} and to heavy nuclei~\cite{Alhassid2008,Ozen2013}. 

\section{SMMC and the calculation of state densities}

\subsection{SMMC method} The Gibbs ensemble $e^{-\beta H}$ describing a nucleus with Hamiltonian $H$ at inverse temperature $\beta$, can be decomposed as a superposition of ensembles $U_\sigma$ of non-interacting nucleons in external auxiliary fields $\sigma(\tau)$ that depend on imaginary time $\tau$ ($0\leq \tau\leq \beta$)
\be\label{HS}
e^{-\beta H} = \int D[\sigma] G_\sigma U_\sigma \;,
\ee
where $G_\sigma$ is a Gaussian weight. This representation is known as the Hubbard-Stratonovich transformation~\cite{HS-trans}. The Hamiltonian $H$ is taken to be a CI shell model Hamiltonian that is defined in a truncated single-particle space with $N_s$ orbitals.  The thermal expectation value of an observable $O$ can be written as 
\be \label{observable}
\langle O\rangle = {\Tr \,(O e^{-\beta H})\over  \Tr\, (e^{-\beta H})} = \frac{\int D[\sigma] W_\sigma \Phi_\sigma \langle O \rangle_\sigma}{\int D[\sigma] W_\sigma \Phi_\sigma} \;,
\ee
where $W_\sigma = G_\sigma |\Tr\, U_\sigma|$
 is a positive-definite function, $\Phi_\sigma = \Tr\, U_\sigma/|\Tr\, U_\sigma|$ is the Monte Carlo sign, and $\langle O \rangle_\sigma = {\rm Tr} \,(O U_\sigma)/ {\rm Tr}\,U_\sigma$.

Since $U_\sigma$ is a one-body propagator, the quantities in the integrands of (\ref{observable}) can be calculated using matrix algebra in the single-particle space. For example, in the grand canonical ensemble
\be
{\rm Tr}\; U_\sigma = \det ( {\bf 1} + {\bf U}_\sigma) \;,
\ee
where ${\bf U}_\sigma$ is the $N_s\times N_s$ matrix that represents $U_\sigma$ in the single-particle space.  The grand canonical expectation value of a one-body observable $O =\sum_{i,j}  O_{ij} a^\dagger_i a_j$ is calculated from
\be\label{1-body}
\langle a_i^\dagger a_j \rangle_\sigma
 = \left[ {1 \over {\bf 1} +{\bf U}^{-1}_\sigma}
\right]_{ji} \;.
\ee
In SMMC we use the canonical ensemble of fixed numbers of protons and neutrons.  This is accomplished by representing the particle-number projection as a discrete Fourier transform~\cite{Ormand1994}. 

The number of auxiliary fields is very large and the integration is carried out by using Monte Carlo methods.  Auxiliary-field samples $\sigma_k$ are chosen according to the positive-definite weight $W_\sigma$ and the expectation value of an observable $O$ in (\ref{observable}) is estimated from
 \be
\langle O\rangle \approx  { {\sum_k  \langle  O \rangle_{\sigma_k} \Phi_{\sigma_k} \over \sum_k \Phi_{\sigma_k}}} \;.
\ee 

\subsection{State density}

In SMMC, we calculate the thermal energy at inverse temperature $\beta$ as the canonical  expectation value of the Hamiltonian, $E(\beta) = \langle H \rangle$. The nuclear partition function $Z(\beta) \equiv \Tr e^{-\beta H}$ is then calculated by integrating the thermodynamic relation  $-{\partial \ln Z / \partial \beta } = E(\beta)$. We have
\be
\ln Z (\beta)= \ln Z (0) - \int_0^\beta E(\beta) d\beta \;.
\ee
where $Z(0)$ is given by the total number of many-particle states in the model space. 
The state density is given by an inverse Laplace transform of  $Z(\beta)$
\be\label{inv-L} 
\rho(E) = {1 \over 2\pi i} \int_{-i \infty}^{i \infty} d\beta \,e^{\beta E} Z(\beta) \;. 
\ee 
We evaluate the average state density in the saddle-point approximation to (\ref{inv-L})~\cite{BM69}
 \be\label{saddle} 
\rho(E) \approx \left(-2 \pi {d E \over d \beta}\right)^{-1/2}  
e^{S(E)}\;, 
\ee 
where $S(E)$ is the canonical entropy given as a Legendre transform of $\ln Z(\beta)$
\be\label{entropy}
S(E) = \ln Z +\beta E \;.
\ee
In (\ref{saddle}) and (\ref{entropy}) we use the value of $\beta$ which is determined by the saddle-point condition 
\be
E = -{\partial \ln Z \over \partial \beta } = E(\beta) \;.
\ee

The density calculated in SMMC is the {\em state} density, in which the magnetic degeneracy $2J+1$ of each level with spin $J$ is taken into account. However, the density measured in the experiments is often the {\em level} density, in which each level is counted once irrespective of its magnetic degeneracy.  We introduced a method~\cite{Alhassid2015,Bonett2013} to calculate the level density $\tilde\rho$ directly in SMMC by using
\begin{eqnarray}
\tilde \rho =  \left\{ \begin{array}{cc} \rho_{M=0}   & \mbox{even-mass nucleus}  \\
\rho_{M=1/2}   &  \mbox{odd mass nucleus} 
\end{array} \right. \;,
\end{eqnarray}
where $\rho_M$ is the density at a given value $M$ of the spin component $J_z$. This $M$-projected density is calculated by implementing a spin-projection method in the Hubbard-Stratonovich transformation~\cite{Alhassid2007}. 

\section{Application to heavy nuclei}

We applied SMMC to nuclei as heavy as the lanthanides using a proton-neutron formalism that allows for different sets of single-particle orbitals for protons and for neutrons. In studies of rare-earth nuclei, we used the $50-82$ shell plus the $1f_{7/2}$ orbital for protons, and the $82-126$ shell plus the $0h_{11/2}, 1g_{9/2}$ orbitals for neutrons. The single-particle energies and orbitals are determined by a Woods-Saxon potential plus spin-orbit interaction.  The interaction includes an attractive monopole pairing interaction and attractive multipole-multipole interactions with quadrupole, octupole and hexadecupole components. The interaction strengths were determined empirically as discussed in Refs.~\cite{Alhassid2008,Ozen2013}.  At large values of $\beta$ the matrices ${\bf U}_\sigma$ become ill-defined and require stabilization. Stabilization methods were developed for strongly correlated electron systems in the grand canonical ensemble~\cite{Loh1992} and we extended them to the canonical ensemble~\cite{Alhassid2008}. 

\subsection{Collectivity in the CI shell model}

Heavy nuclei are known to exhibit various types of collectivity that are well described by empirical models. An important question is whether these types of collectivity can be described within a spherical CI shell model framework, in which the single-particle model space is truncated. The large dimensionality of the many-particle model space in heavy nuclei necessitates the use of quantum Monte Carlo methods such as SMMC. The various types of collectivity are usually identified by their characteristic spectra. While SMMC is a powerful method that allows the accurate calculation of thermal observables, it is difficult to extract detailed spectroscopic information from the thermal expectation values (\ref{observable}). 

To overcome this difficulty, we identified an observable whose low-temperature  behavior is sensitive to the type of collectivity.  This observable is $\langle {\bf J}^2\rangle_T$, where ${\bf J}$ is the total angular momentum of the nucleus. Assuming an even-even nucleus which is either vibrational or rotational, the low-temperature behavior of  $\langle {\bf J}^2\rangle_T$ is given by
\begin{eqnarray}\label{J2-theory}
\langle \mathbf{J}^2 \rangle_T \approx
 \left\{ \begin{array}{cc}
 30 { e^{-E_{2^+}/T} \over \left(1-e^{- E_{2^+}/T}\right)^2} & \mbox{vibrational band}  \\
 \frac{6}{E_{2^+}} T & \mbox{rotational band}
 \end{array} \right. \;,
\end{eqnarray}
where $E_{2^+}$ is the excitation energy of the lowest $2^+$ level. 

\begin{figure}[t]
\begin{center}
\includegraphics[width=16cm]{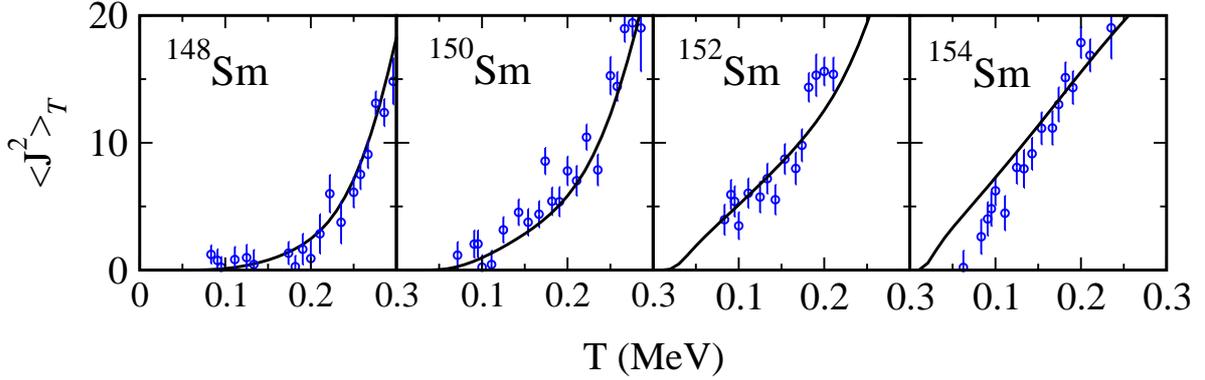}
\caption{ $\langle {\bf J}^2\rangle_T$ vs.~temperature $T$ in the even samarium isotopes $^{148-154}$Sm. The SMMC results (open circles) are compared with results deduced from experimental data (solid lines; see text). Adapted from Ref.~\cite{Ozen2013}.}
\label{J2}
\end{center}
\end{figure}

In Fig.~\ref{J2}, we show the SMMC results (open circles) for $\langle {\bf J}^2\rangle_T$ as a function of temperature $T$ for a family of even samarium isotopes $^{148-154}$Sm.  This family of isotopes are known to describe a crossover from vibrational collectivity in the spherical $^{148}$Sm nucleus to rotational collectivity in the deformed $^{154}$Sm nucleus. We observe in the SMMC results for $\langle {\bf J}^2\rangle_T$  a crossover from a ``soft'' response to temperature in $^{148}$Sm to a rigid linear response in $^{154}$Sm, in agreement with (\ref{J2-theory}). 

The solid lines in Fig.~\ref{J2} are obtained from the experimental data by taking into account a complete set of measured energy levels up to certain excitation energy and by using a back-shifted Bethe formula (BBF)  above that energy. This density is determined from level counting at low energies and neutron resonance data at the neutron separation energy.

\subsection{State densities}

We calculated the SMMC state densities in families of samarium and neodymium isotopes~\cite{Ozen2013,Ozen2015}. To compare with experimental data, it is necessary to determine the excitation energy $E_x=E - E_0$, where $E_0$ is the ground-state energy.  It is thus important to determine an accurate ground-state energy. 

\subsubsection{Even-even nuclei}

\begin{figure}[htb]
\begin{center}
\includegraphics[width=16cm]{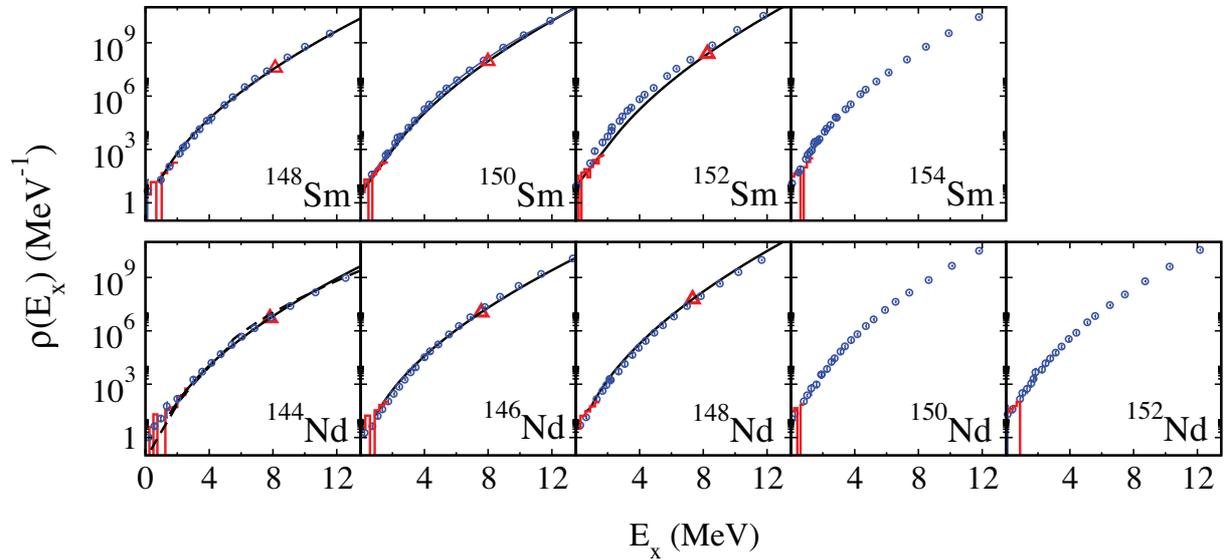}
\caption{State densities vs.~excitation energy $E_x$ in even-mass samarium (top panels) and neodymium (bottom panels) isotopes. The SMMC densities (open circles) are compared with level counting data (histograms) and neutron resonance data (triangles) when available~\cite{Ripl3}.  The neutron resonance data was converted to a total state density assuming a spin cutoff model with rigid-body moment of inertia. The solid lines are empirical BBF densities (see text). Adapted from Refs.~\cite{Ozen2013,Alhassid2014a}.}
\label{rho_even}
\end{center}
\end{figure}

The ground-state energy $E_0$ of even-even nuclei can be determine accurately from large $\beta$ calculation of the thermal energy.  Fig.~\ref{rho_even} shows the SMMC state densities (open circles) in even-mass samarium and neodymium isotopes. The results are generally in good agreement with level counting data at low excitation energies (histograms) and with neutron resonance data (triangles) when available~\cite{Ripl3}. The solid lines describe BBF state densities determined empirically from the level counting and the neutron resonance data. 

\subsubsection{Odd-even nuclei}

\begin{figure}[htb]
\begin{center}
\includegraphics[width=16cm]{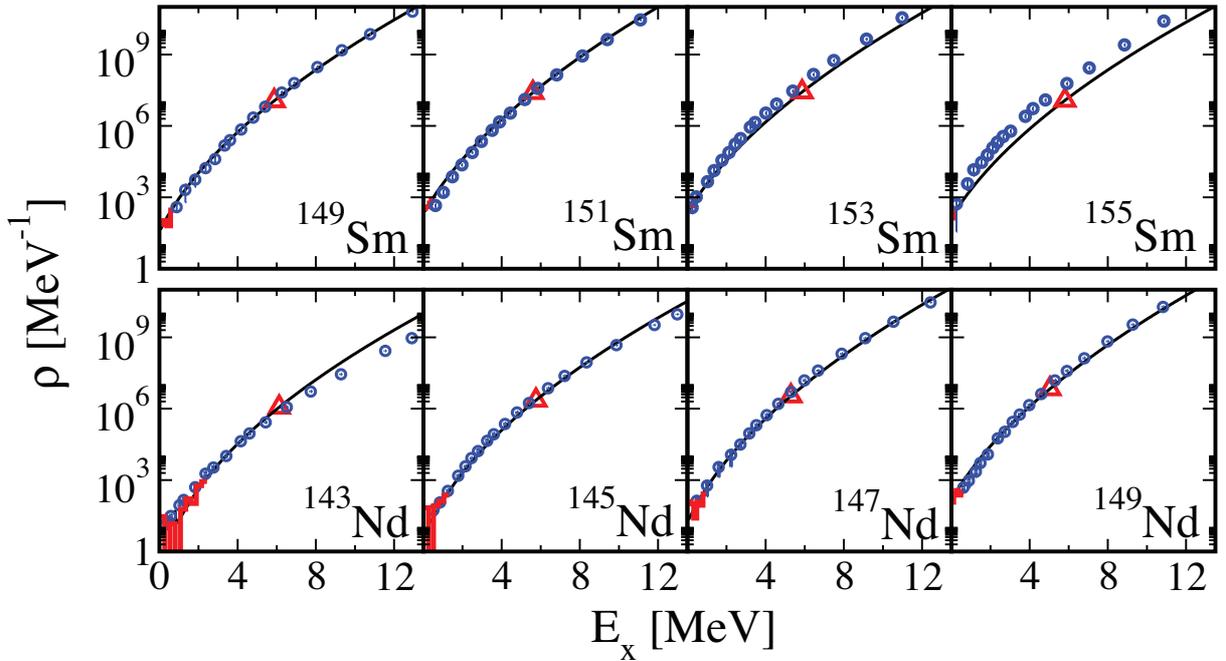}
\caption{State densities vs.~excitation energy $E_x$ in odd-mass samarium (top panels) and neodymium (bottom panels) isotopes. Symbols and lines are as in Fig.~\ref{rho_even}. Adapted from Ref.~\cite{Ozen2015}.}
\label{rho_odd}
\end{center}
\end{figure}

SMMC calculations at low temperatures in odd-even nuclei have a sign problem that originates from the projection on odd number of particles, leading to large error bars in the thermal energy. Consequently, we can calculate the thermal energy only up to $\beta \sim 4$ MeV$^{-1}$ and it is difficult to determine an accurate ground-state energy.  A method to calculate an accurate ground-state energy for a system with odd number of particles despite the odd particle-number  sign problem was introduced in Ref.~\cite{Mukherjee2012} and applied to mid-mass nuclei. However, its application in heavy nuclei requires additional development.  For the heavy nuclei we determine the ground-state energy $E_0$ from a one-parameter fit of the SMMC thermal energy to the thermal energy calculated from experimental data~\cite{Ozen2015}. The corresponding state densities in odd-mass samarium and neodymium isotopes are shown in Fig.~\ref{rho_odd} and compared with experimental data.

\section{Deformation in the CI shell model}

Knowledge of state densities as a function of nuclear deformation is useful in the modeling of fission. Nuclear deformation is a key concept in our understanding of the physics of heavy nuclei. However, it is based on a mean-field approximation that breaks rotational invariance. The challenge is to study nuclear deformation in the framework of the CI shell model approach which preserves rotational symmetry. 

\subsection{Quadrupole distributions in the laboratory frame}

We calculated the SMMC distribution $P_T(q)$ of the axial quadrupole operator $\hat Q_{20}$  in the laboratory frame~\cite{Alhassid2014} at inverse temperature $\beta$ by using its Fourier representation 
\be
\label{prob}
P_T(q)={1\over {\rm Tr}\,  e^{-\beta H}} \int_{-\infty}^\infty
{d \varphi \over 2 \pi} e^{-i \varphi q }\, {\rm Tr}\, \left(e^{i \varphi Q_{20}} e^{-\beta
 H} \right)
\ee
together with the Hubbard-Stratonovich representation (\ref{HS}) of $e^{-\beta H}$.  We divide an interval $[-q_{\rm max},q_{\rm max}]$  into $2M+1$ intervals of length $\Delta q=2q_{\rm max}/(2M+1)$ and use a discrete Fourier repsrentation
\begin{equation}\label{fourier-q}
{\rm Tr}\left(\delta(Q_{20} - q_m) U_\sigma \right) \!\!  \approx \!\! {1\over 2 q_{\rm max}}\! \! \sum_{k=-M}^M
  \!\!\!  e^{-i \varphi_k q_m} {\rm Tr}(e^{i \varphi_k Q_{20}} U_\sigma) \;,
\end{equation}
where $q_m=m \Delta q$ ($m=-M,\ldots,M$) and $\varphi_k = \pi k/q_{\rm max}$ ($k=-M,\ldots, M$).

\begin{figure}[htb]
\begin{center}
\includegraphics[width=16cm]{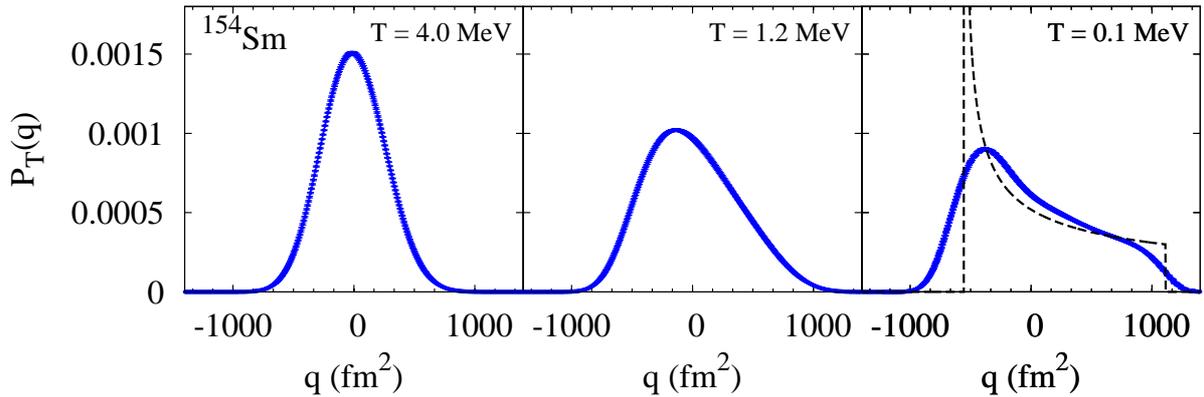}
\caption{The SMMC quadrupole distributions $P_T(q)$ in the laboratory frame for $^{154}$Sm at  a low temperature $T=0.1$ MeV, an intermediate temperature $T=1.2$ MeV (close to the HFB shape transition temperature) and a high temperature $T=4$ MeV. Adapted from Ref.~\cite{Alhassid2014}.}
\label{Sm154_zvq}
\end{center}
\end{figure}

The distributions $P_T(q)$ are shown for $^{154}$Sm in Fig.~\ref{Sm154_zvq} at three temperatures. At a low temperature ($T=0.1$ MeV) we find a skewed distribution that is qualitatively similar to that of a prolate rigid rotor (dashed line) whose intrinsic axial quadrupole moment is taken from a finite-temperature Hartree-Fock-Bogoliubov (HFB) approximation. In the HFB approximation, we observe a shape transition from a deformed to a spherical shape around a temperature of $T=1.2$ MeV. At this temperature the quadrupole distribution is still skewed. At high temperatures, e.g., $T=4$ MeV, the distribution is close to a Gaussian. 
We also calculated the quadrupole distributions for $^{148}$Sm which is spherical in its mean-field ground state and found that they are close to a Gaussian already at low temperatures. We conclude that the quadrupole distribution in the laboratory frame is a clear model-independent signature of nuclear deformation in the framework of the rotational-invariant CI shell model approach.  

\subsection{Quadrupole distributions in the intrinsic frame}

For the formalism developed here to be useful for calculating level densities as a function of intrinsic deformation, it is necessary to determine the quadrupole distribution in the intrinsic frame.  The intrinsic frame is usually defined within a mean-field approximation, so a direct calculation of this distribution in the CI shell model approach is not feasible. To overcome this difficulty, we consider the second rank quadrupole tensor $q_{2\mu}$ ($\mu=-2,\ldots, 2$). In the intrinsic frame (characterized by a set of three Euler angles $\Omega$), the quadrupole shape is characterized by two shape parameters $\beta,\gamma$.  Note that we use the same symbol $\beta$ to denote both the inverse temperature and the axial shape parameter, and the correct meaning should be clear from the context. The probability density $P_T(\beta,\gamma)$ in the intrinsic frame is a rotational invariant and therefore we can expand $-\ln P_T(\beta,\gamma)$ in rotational invariant combinations of $q_{2\mu}$. There are only three quadrupole invariants up to fourth order in $q_{2\mu}$ and they are given by $\beta^2, \beta^3\cos 3\gamma$ and $\beta^4$.  Thus 
\be\label{landau}
-\ln P_T(\beta,\gamma) = N + A\beta^2 -B \beta^3\cos 3\gamma + C \beta^4 + \ldots \;,
\ee
where $A,B, C$ are temperature-dependent parameters and $N$ is a normalization constant. Eq.~(\ref{landau}) is similar to the Landau expansion of the free energy in which the quadrupole tensor is considered as the order parameter of the shape transition~\cite{Levit1984,Alhassid1986}. The parameters $A,B,C$  in (\ref{landau}) can be determined from the expectation values of the above three quadrupole invariants. In calculating these expectation values from the density $P_T$ it is necessary to take into account the volume element  
\be
\prod_\mu d q_{2\mu} \propto \beta^4 |\sin 3\gamma | \, d\beta \, d\gamma\, d\Omega \;.
\ee
In Ref.~\citeonline{Alhassid2014} we showed that the above three quadrupole invariants are related to moments of $Q_{20}$ in the laboratory frame and thus can be directly calculated  from the SMMC distribution $P_T(q)$ in the laboratory frame.  

\begin{figure}[htb]
\begin{center}
\includegraphics[width=16cm]{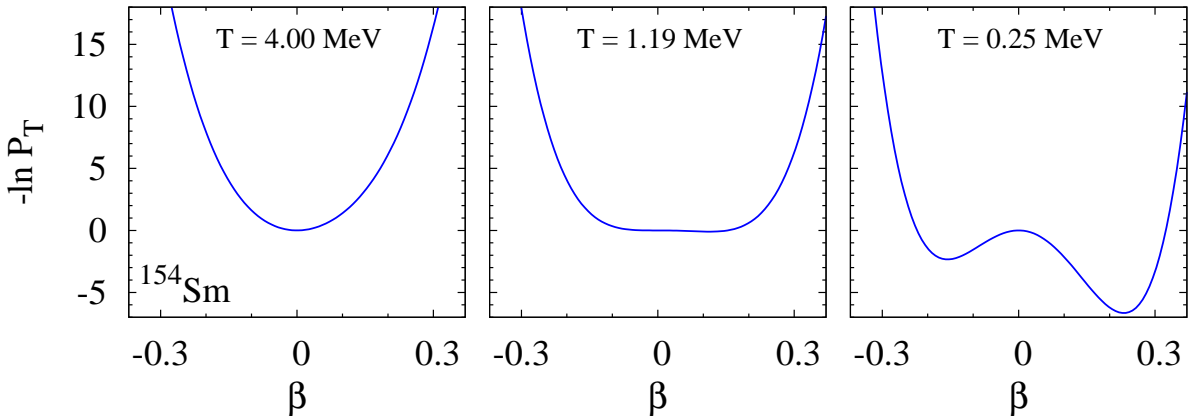}
\caption{$-\ln P_T(\beta,\gamma=0)$ vs.~axial deformation $\beta$ for $^{154}$Sm at three temperatures.}
\label{Sm154_logP}
\end{center}
\end{figure}

In Fig.~\ref{Sm154_logP} we show $-\ln P_T(\beta,\gamma=0)$ as a function of the axial deformation parameter $\beta$ for $^{154}$Sm at three temperatures $T=0.25, 1.19$ and $4$ MeV. The curves in Fig.~\ref{Sm154_logP}, derived in the CI shell model framework without the use of a mean-field approximation, seem to mimic the shape transition that is found in the HFB mean-field approximation. The minima of these curves describe a shape transition from a prolate minimum at low temperatures  (e.g., $T=0.25$ MeV) to a spherical  minimum at high temperatures (e.g., $T=4$ MeV). 

By using the saddle-point approximation, the distributions $P_T(\beta,\gamma)$ at constant temperature can be converted to intrinsic shape distributions $P_{E_x}(\beta,\gamma)$ at constant excitation energy. The joint level density distribution as a function of excitation energy and intrinsic deformation can then be determined from $\rho(E_x,\beta,\gamma)= \rho(E_x) P_{E_x}(\beta,\gamma)$.

\section{Conclusion}
The SMMC method is a powerful method to calculate microscopically nuclear state densities in the presence of correlations, and was applied to nuclei as heavy as the lanthanides. We have also introduced a method to calculate the distribution of the quadrupole deformation in both the laboratory frame and in the intrinsic frame within the rotationally invariant CI shell model approach.  We plan to use this method to calculate level densities as a function of intrinsic deformation. 

\section*{Acknowledgements}
This work was supported in part by the DOE grant Nos.~DE-FG-0291-ER-40608 and DE-FG02-00ER411132, by Grant-in-Aid for Scientific Research (C) No. 25400245 by the JSPS, Japan, and by the Turkish Science and Research Council (T\"{U}B\.{I}TAK) grant No. ARDEB-1001-110R004 and ARDEB-1001-112T973. The research presented here used resources of the National Energy Research Scientific Computing Center, which is supported by the Office of Science of the U.S. Department of Energy under Contract No.~DE-AC02-05CH11231. It also used resources provided by the facilities of the Yale University Faculty of Arts and Sciences High Performance Computing Center.

\end{document}